\documentstyle[aps,prb]{revtex}
\begin{document}
\maketitle
\twocolumn[]

{\Large\bf The Quantized Hall Insulator: A New Insulator 
in Two-Dimensions}

{\bf  M. Hilke, D. Shahar$^1$, S.H. Song$^2$, D.C. Tsui, 
Y.H. Xie$^3$ and Don Monroe$^3$}

{\em 
Dpt. of Electrical Engineering, Princeton University, Princeton, New 
Jersey, 08544}

{\em $^1$ Present address: Dept. of Condensed Matter Physics, Weizmann 
Institute, 
Rehovot 76100, Israel}

{\em $^2$ Present address: LG Semicon, Choongbuk, 361-480, Korea}

{\em $^3$  
Bell Laboratories, Lucent Technologies, Murray Hill,  New 
Jersey, 07974}

{\bf 
Quite generally,
an insulator is theoretically defined by a vanishing conductivity tensor 
at the absolute zero of temperature {\em T}.
In classical insulators, such as band insulators, vanishing 
conductivities 
lead to diverging resistivities.
In other insulators, in particular when a high magnetic field {\em B} is 
added, 
it is possible that 
while the magneto-resistance, $\rho_{xx}$, diverges, the Hall 
resistance, $\rho_{xy}$, remains finite, which is known as a 
Hall insulator (HI)\cite{KLZ}.
In this letter we demonstrate experimentally the existence of another, 
more exotic, insulator. This insulator, which terminates the quantum 
Hall effect (QHE) series in a two-dimensional electron system, is 
characterized by a $\rho_{xy}$ 
which is approximately {\em quantized} in the quantum unit of resistance 
$h/e^2$.  This insulator is termed a quantized Hall insulator 
(QHI)\cite{efrat}.
In addition we show that for the same sample, the insulating state 
preceding the QHE series, at low-$B$, is of the HI kind.
}

Experimentally the identification 
of an insulating phase is based on extrapolating the  measured 
$\rho_{xx}(T)$ at finite {\em T} to $T=0$. This is always an ambiguous 
process. 
However, when $\rho_{xx}$ is exponentially increasing as $T\rightarrow 
0$, 
the state is usually considered to be an insulator. Unfortunately, the 
divergent $\rho_{xx}$ seriously hinders the determination of $\rho_{xy}$, 
since even small Hall-contact misalignment will result in a large 
overriding 
signal from the diverging $\rho_{xx}$. It is possible, to a certain 
degree, to circumvent this difficulty by symmetrizing the measurement. 
This can be achieved by reversing the 
{\em B}-field  orientation, as the contribution of $\rho_{xx}$ is 
symmetric in {\em B} as opposed to antisymmetric for $\rho_{xy}$. 
The effectiveness of this 
procedure is demonstrated  in the inset of Fig. 1, where the Hall 
resistances obtained for the two opposite {\em B}-field directions are in 
dotted lines and the average, i.e., $\rho_{xy}$ is presented with a  
solid line. For the remainder of this letter all  
Hall resistivities are obtained using this method.

We now turn to discuss our results, where our first task is to identify 
the different phases.
The transition between insulating and quantum Hall 
phases can be characterized by a critical {\em B}-field value, for which 
$\rho_{xx}$ is {\em T}-independent and where the derivative of the 
{\em T}-dependence changes sign on  
each side of the transition. 
By plotting $\rho_{xx}$ at two different 
{\em T} 's, we can therefore extract the transition points. 
In Fig. 1 we have plotted $\rho_{xy}$ together with $\rho_{xx}$
as a function of $B$. With increasing {\em B}, 
transition points at $B=2.2$ T and  at $B=B_{C}=6.06$ T,  
can be 
identified from the crossing of the two $\rho_{xx}$ curves obtained 
at different {\em T} 's. In between these transitions we have the usual 
{\em QH} state, which is bordered on both sides by insulators.  
For clarity $\rho_{xx}$ is normalized  
to $\rho_c=\rho_{xx}(B_{C})$.\cite{shahar}

\input epsf
\begin{figure}
\epsfysize=7cm

\epsfbox{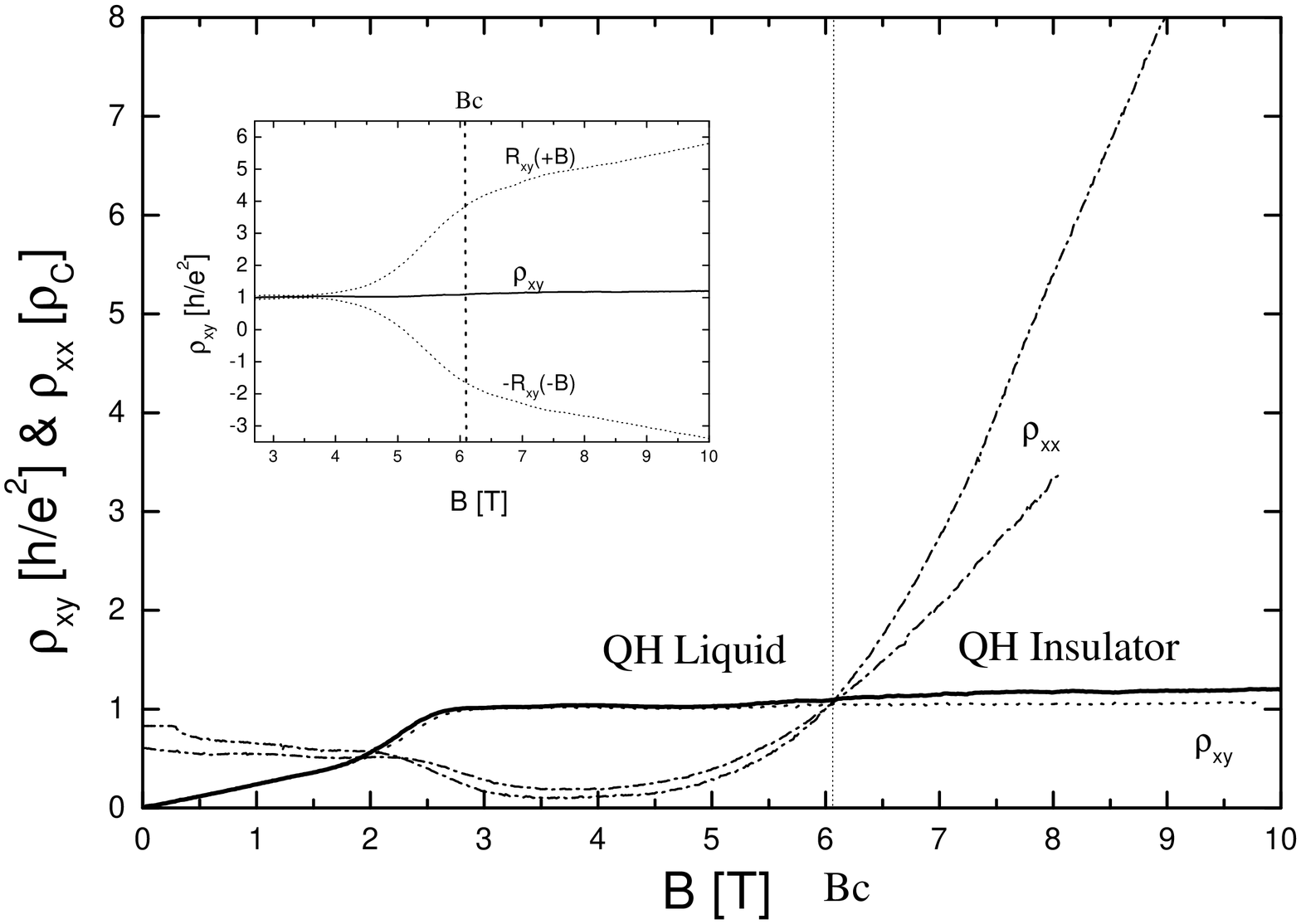}

\caption{
The Hall and diagonal resistivities as a function of 
{\em B}-field. 
The solid line is the Hall resistivity measured at 
T=300 mK and a current I=200 nA, whereas the dotted line 
is for I=400 nA. 
The dash-dotted lines are $\rho_{xx}$ at T=1.2 K (the 
uppermost curve) and at T=4.2 K (the lower curve). $V_G$=5.2 V, 
$\rho_c=1.65 h/e^2$ and $B_C$=6.06 T.
The inset shows the Hall resistances for $+B$ and $-B$ in dotted lines 
and $\rho_{xy}$ with a solid line. The experimental error is mainly given 
by 
$\left|\frac{\Delta R_{xy}}{R_{xy}}\right|\cdot\left( 1+ 
\frac{(\rho_{xy}-R_{xy}(+B))^2}{\rho^2_{xy}}\right)^{1/2} $. 
The first term can 
be estimated from the fluctuations around $\nu=1$, i.e., $\simeq 4\%$ and 
the second term is $\simeq 5$, leading to a total inaccuracy of 20 \%. With 
increased current the first term is reduced to $\simeq 3\%$ and the second 
to $\simeq 2$, yielding a total inaccuracy of 6 \%.
}
\end{figure}

Focusing first on the high {\em B}-field region, at $B>B_{C}$, a 
striking 
observation that can be made in Fig. 1, is the large range 
over which $\rho_{xy}$ remains nearly constant and 
 close to its quantized value 
of $h/e^2$. Indeed, between 2.7 T and 10 T, the deviation of $\rho_{xy}$
from $h/e^2$ is less than 20 $\%$. When doubling the current  
the symmetric contribution to $\rho_{xy}$ is further
reduced and 
therefore the deviation is much smaller, i.e., less than 5 $\%$.
This could be due to the good contact alignment as higher currents tend to 
render the system more homogeneous. 
In terms of the Landau level filling factor $\nu$, 
this means that 
$\rho_{xy}$  remains approximately quantized between 
$\nu=1.5$ and $\nu=0.4$. The 
transition point is at
$\nu_c=0.75$.

As shown in Fig. 1, a 5-20\% accuracy was obtained. In order to increase 
this accuracy even further but without increasing the current and 
to analyze the 
{\em T}-dependence,  
we used a standard low frequency lock-in technique, with 1 nA currents. 
In the inset of Fig. 2, we have extracted the {\em T}-dependence of  
$\rho_{xy}$ from the main figure, where $\rho_{xx}$ and $\rho_{xy}$ are 
plotted as a function of $B$.
The important result here, is the clear indication that $\rho_{xy}$ 
saturates towards the quantized value, within 2$\%$, 
at low enough temperatures (below 2 K).

\input epsf
\begin{figure}
\epsfysize=7cm

\epsfbox{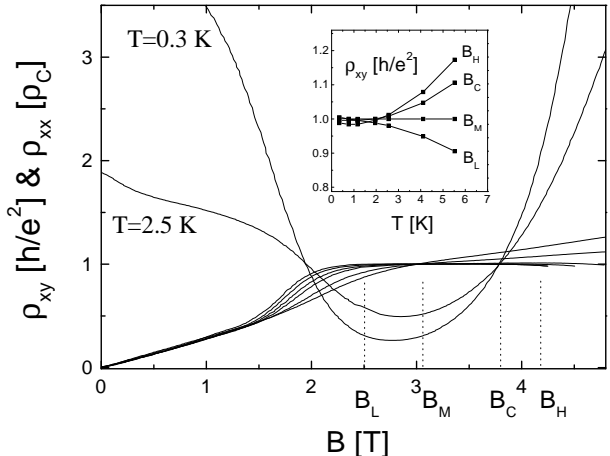}


\caption{
The Hall and diagonal resistivities as a function 
of B-field for different temperatures. The T's are 0.3, 1.2, 
2, 2.7, 4.2  and 5.5 K. $V_G$=5.4 V, $\rho_c=1.73h/e^2$ and $B_C=3.8$ T.
In the inset we have plotted the {\em T}-dependence of $\rho_{xy}$.
The constant curve in the inset corresponds to the 
{\em T}-independent $\rho_{xy}$ at $B_M=3.1$ T, which is close to $\nu=1$.
$B_L$=2.5 T corresponds to 
a typical {\em B}-field $\nu>1$, where the Hall resistance decreases 
with {\em T}. $B_C$=3.8 T corresponds to the critical value 
defined by the {\em T}-independent $\rho_{xx}$ and 
$B_H$=4.2 T to a typical {\em B}-field larger than $B_C$.}
\end{figure}

In addition we were also interested in studying 
$\rho_{xy}$ when $\rho_{xx}$ is highly insulating even at low 
{\em B}-field, i.e., below the low {\em B}-field transition.
Fig. 2 was therefore 
obtained by increasing the effective disorder, i.e., we  reduced 
the density of the sample by applying a higher voltage ($V_G$) on the front gate 
of our 2D hole system (2DHS) confined in a Ge/SiGe quantum well.
A more detailed description 
of the sample (X335) and its properties are given in ref. \cite{hilke}.
The low 
{\em B}-field transition 
is observed at $B=2.2$ T in Fig. 1 and at $B=1.9$ T in Fig. 2, where we 
have applied a higher $V_G$. 
For fields smaller than the value at the low {\em B}-field 
transition, the system is insulating and  
$\rho_{xy}$ follows the classical dependence 
on the magnetic field, i.e., $\rho_{xy}\simeq B/nec$. We observe no 
temperature dependence of $\rho_{xy}$ for low enough fields. At the same 
time $\rho_{xx}\simeq 7h/e^2$, at the lowest measured {\em T}, indicating that 
we are in the presence of a strong insulator, more precisely a {\em HI}.
In order to measure only the diffusive part 
we limited the value of the maximum phase shift to be $2^o$.   
Finally, we would like to mention that we also observed this 
quantization (within experimental accuracy) 
when we used the current contacts as Hall probes and the Hall 
contacts as current leads, hence a mesoscopic effect can be ruled out. 
The same results were obtained using a sample from a different wafer (X334), 
having a very similar structure.

Most previous experiments  in the fractional 
quantum Hall regime \cite{bob,goldman,sajoto,goldman2} 
suggested a Hall resistance following the classical
expression $\rho_{xy}\simeq B/nec$. The main difference between these samples 
and 
ours is their much higher mobilities, which can hence account for the 
different behavior. In the IQH regime but with higher mobilities than 
in our samples  
Shahar et al. 
\cite{shahar4,shahar2}
observed that the Hall resistivity remains close to the 
quantized  value around the transition, but at their lowest 
temperatures, they observed a small deviation from the quantized value. 
The origin of this deviation is most likely $\rho_{xx}$ mixing, which 
becomes too large at very
low {\em T} and cannot be removed by symmetrizing. In our case, however,  
we managed to minimize this effect, which translates into an 
even better quantization at low {\em T}.
In previous experiments, with samples of lower mobilities, $\rho_{xy}$ either 
had a 
strong {\em T}-dependence \cite{hughes} or a low accuracy \cite{kravchenko} 
and others resolved only the $\nu=2$ to insulator transition 
\cite{hughes,jiang,lee} along with a 
small {\em B}-field range 
in the insulating 
phase and therefore no generic behavior was demonstrated. 
In this work we presented results where we observe a 
Hall resistivity equal to the quantized value $h/e^2$, within 
experimental accuracy, as deep as 4 T beyond the transition point $B_C=6.06$ T.

While most theories agree on a quantized value of $\rho_{xy}$ at the 
transition, discrepancies exist for the value inside the insulator. 
First let us point out where the difficulties come from. In standard 
transport theories, such as in Boltzmann's linear response theory or 
in the Kubo formalism, one usually calculates conductivities. Using matrix 
inversion (in 2D) one obtains $\rho_{xy}=\frac{\sigma_{xx}}{\sigma_{xx}^2+
\sigma_{xy}^2}$. As in an insulator the conductivities vanish, any small 
correction to the conductivities can alter the value of $\rho_{xy}$ in the 
limit $\sigma\rightarrow 0$.
In the Kubo formulation, the conductivities are expressed as a function 
of the frequency $\omega$ and $T$. The physically relevant approach, for 
obtaining the diffusive part of the transport coefficients, is to first 
take the limit $\omega\rightarrow 0$ and then $T\rightarrow 0$.  In the 
opposite limit only the reactive 
part $\sigma_{xx}\sim i\omega$ is obtained \cite{efetov,zhang}, 
leading to a finite 
$\rho_{xy}$, for any type of insulator. In most cases the two limits do not
commute.  

In the 
framework of Kivelson, Lee and Zhang's (KLZ) global phase diagram \cite{KLZ}, 
the quantum Hall liquid phases 
are bordered by an insulator. 
In the insulating phase KLZ obtained 
a finite $\rho_{xy}$ in a RPA-type of approximation, by evaluating the 
diffusive part. Concentrating on the transition region, Shimshoni and others 
\cite{sondi,duality}, considered the duality/particle-hole symmetry, 
leading to a quantized $\rho_{xy}$ in the critical regime. In this 
context the experimental finding of an approximate quantization, throughout 
the transition, confirms the large region of validity of this duality. When 
following KLZ's analysis and considering the insulating phase 
outside the critical region, an exact quantization 
would imply the existence of an additional phase transition, i.e., 
from a {\em QHI} to {\em HI} behavior. This would certainly be a very 
interesting search to pursue in future experiments.

In a different type of approach
a  semi-classical network model was considered, yielding a 
{\em QHI} \cite{efrat,pryadko}. 
This network model describes the insulating phase, 
where {\em QH}-liquid states exist 
in long range random potential minima.
Transport is then obtained 
via a scattering matrix, parameterizing the tunneling 
from one {\em QH}-liquid to the other. For the case
where phase coherence is destroyed after each scattering event, i.e., in 
the semi-classical limit, $\rho_{xy}$ was shown to remain quantized.

Our main results can be  stated as follows: at very low 
temperatures and high {\em B}-fields, $\rho_{xy}\simeq h/e^2$  
deep in the insulating phase, hence 
forming a {\em QHI}. 
At slightly higher temperatures $\rho_{xy}$ starts 
to deviate from its quantized value and approaches its classical 
value proportional to $B$. At low {\em B}-fields 
$\rho_{xy}$ follows its classical value even in the strongly insulating 
case, 
i.e., forming a {\em HI} state.
In the next paragraph we describe an interesting 
consequence of this quantization.

Recently Dykhne and Ruzin \cite{ruzin} suggested that the conductivities 
follow a semicircle relation, where each semicircle formed 
by $\sigma_{xx}$ as a function of $\sigma_{xy}$, is centered 
around $\sigma_{xy}=(n+1/2)e^2/h$, where $n$ is an integer. 
We wanted to test this relation for
the lowest Landau level, i.e., the semicircle centered at $e^2/2h$. 
Fig. 3 presents the result, showing a clear evidence 
for this relation.

\begin{figure}
\epsfysize=6cm
\begin{center}
\leavevmode
\epsfbox{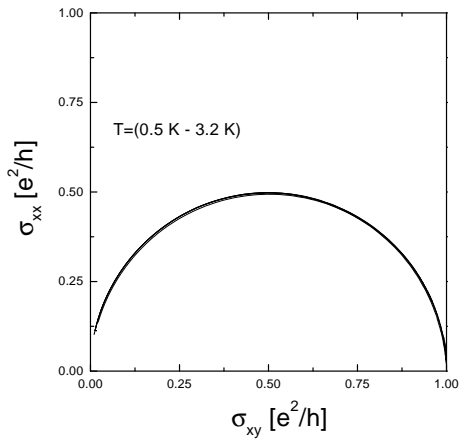}
\end{center}

\caption{
$\sigma_{xx}$  as a function of $\sigma_{xy}$ for 
different temperatures. The {\em T} 's are 0.5, 0.8, 
1.2, 1.8 and 3.2 K. $V_G=5.2$ V and I=1 nA.
}
\end{figure}

This result is in fact not very surprising taking into account 
that we have a quantized Hall resistance deep inside the 
insulator. Indeed, 
\begin{equation}
\sigma_{xx}^2+\left(\sigma_{xy}-\frac{e^2}{2h}\right)^2=
\left(\frac{e^2}{2h}\right)^2+\frac{1-(e^2/h)\rho_{xy}}
{\rho_{xx}^2+\rho_{xy}^2}, 
\end{equation}
simply using matrix inversion. As noted by Shahar et al. \cite{shahar2}, 
the semicircle relation follows 
directly 
as long as $\rho_{xy}=h/e^2$. In addition it is 
completely independent of the value of $\rho_{xx}$. We emphasize that 
this 
relation holds for our entire range of {\em T}.

Ando \cite{ando} first calculated numerically the dependence of 
$\sigma_{xx}$ 
on $\sigma_{xy}$ and found a similar relation to the semicircle function.
In the self consistent Born approximation  and scaling theories a 
behavior of this kind 
is also expected \cite{pruisken}. A recent numerical study by R.N. Bhatt and 
S. Subramanian (private 
communication), using symmetric and non-symmetric random potentials 
in a lowest Landau level model, confirmed this result.
For higher Landau levels Wei et al. \cite{wei} and others \cite{shahar2} 
showed similar dependences. 
This is however the first clear demonstration of this relation for 
$\nu\leq 1$ and $\sigma_{xx}$ and $\sigma_{xy}$ both smaller than 
$e^2/2h$.

Summarizing, we have demonstrated experimentally the existence of 
a {\em HI} and a {\em QHI}. The {\em QHI} leads interestingly to the 
semicircle relation. The experiments presented covered a certain 
range of parameters, in particular in temperature and in disorder. 
Different behaviors occurring at even lower 
temperatures or higher mobilities are therefore possible and would 
be interesting to investigate in future work. 
The existence of these insulators, however, can serve as guidelines for 
a more complete understanding of two-dimensional systems in strong 
magnetic 
fields.

{\bf Acknowledgements:} We thank A. Auerbach, L.P. Pryadko, E. Shimshoni and 
S.L. Sondhi for discussions. This work was supported in part by the National 
Science Foundation and MH was supported by the Swiss National Science 
Foundation.

\vskip.2cm

Correspondence should be addressed to M.H. (e-mail: Hilke@ee.princeton.edu).

\end{document}